\documentclass[11pt]{article}

\usepackage[utf8]{inputenc}
\usepackage[T1]{fontenc}
\usepackage{mathptmx}            
\usepackage[scaled=0.92]{helvet} 
\usepackage{geometry}
\usepackage{microtype}
\usepackage{amsmath,amssymb}
\usepackage{graphicx}
\usepackage{booktabs}
\usepackage{array}
\usepackage{xcolor}
\definecolor{linkblue}{RGB}{20,78,121}
\usepackage{caption}
\usepackage{titlesec}
\titleformat{\section}{\normalfont\sffamily\large\bfseries}{\thesection}{0.6em}{}
\titleformat{\subsection}{\normalfont\sffamily\normalsize\bfseries}{\thesubsection}{0.6em}{}
\titlespacing*{\section}{0pt}{1.4ex plus .2ex}{0.8ex}
\usepackage{authblk}

\usepackage[numbers,sort&compress]{natbib}
\usepackage{fancyhdr}
\usepackage[colorlinks=true,allcolors=linkblue,breaklinks=true]{hyperref}

\fancypagestyle{plain}{\fancyhf{}}

\title{\sffamily\bfseries The Audit Gap in Blockchain Security:\\[2pt]
A Four-Year Empirical Study of Public Audit Findings\\ and Real-World Exploit Incidents}

\author[1]{Stefan Beyer}
\affil[1]{Oak Security \quad \texttt{stefan@oaksecurity.io}}
\date{April 2026}

\begin{document}
\maketitle
\thispagestyle{plain}

\begin{abstract}
\noindent
This paper presents an empirical analysis of the Web3 security landscape over the four-year and three-month period from 1 January 2022 to 27 March 2026. The dataset combines 23{,}818 public audit findings produced by 22 independent security firms with 218 real-world exploit incidents documented by \emph{rekt.news}, representing aggregate losses of approximately US\$7.76 billion. We report three central findings. First, the distribution of audit findings---by severity, category, and technology stack---is substantially stable across the observation window, with the Critical-plus-High share remaining within a 15--17\% band in every complete year. Second, the categorical distribution of realised exploit losses does not correspond to the categorical distribution of audit findings: private-key compromise, phishing, and social-engineering vectors account for approximately 49.6\% of cumulative losses yet represent a negligible share of published audit findings. Third, realised losses exhibit extreme concentration: the eight largest incidents account for 50.6\% of cumulative dollar losses and the twenty largest for 71.4\%, a distributional shape inconsistent with Gaussian assumptions. Throughout, we adopt the analytical convention that audit outputs and exploit outputs describe different populations and present the two datasets in parallel rather than as directly comparable samples.
\end{abstract}

\vspace{0.5em}

\vspace{1.0em}

\section{Introduction}
\label{sec:intro}
Empirical study of the Web3 security domain is hindered by the distributed and heterogeneous nature of its evidentiary base. Audit reports are produced by dozens of firms, in varying formats, published through both repository-based and web-native channels, and they differ substantively in how findings are classified. Exploit data is similarly fragmented, tracked primarily by a small number of journalistic and research publications, and subject to disagreement about what should count as a distinct incident, how to value non-liquid assets, and how to attribute root cause.

This paper attempts a consolidated view across both sides of that evidentiary base over a four-year and three-month observation window. The motivating question is simple: across all public audit output and all documented real-world incidents, do the patterns that auditors surface correspond to the patterns that result in realised losses, and if not, in what ways do they differ? The analysis is conducted at the aggregate level; no single audit report or exploit is examined in isolation.

\paragraph{Related work.}
Prior systematisations of Web3 security have largely been taxonomic or mechanism-focused. Atzei et al.~\citep{atzei2017survey} provide an early survey of attacks on Ethereum smart contracts, organising vulnerabilities by language- and blockchain-level cause. Werner et al.~\citep{werner2022sok} survey the DeFi protocol stack and its security assumptions, and Zhou et al.~\citep{zhou2023sok} catalogue DeFi attack techniques and incidents, reporting---consistent with the present study---that price-oracle manipulation and permissionless-interaction attacks are among the most frequent incident types yet receive disproportionately little academic attention.

A second line of work bears directly on the gap this paper measures: the weak coupling between vulnerabilities that are \emph{detectable} and those that are \emph{exploited}. Perez and Livshits~\citep{perez2021vulnerable} show that of 23{,}327 contracts flagged as vulnerable by six analysis tools, fewer than 2\% were ever exploited, and that the funds genuinely at risk were concentrated in a small number of contracts. Chaliasos et al.~\citep{chaliasos2024tools} evaluate five widely used security tools against 127 high-impact real-world attacks and survey practitioners, finding that the tools would have prevented only a minority of the attacks and that the most damaging incidents fall outside their detection scope. Domain-specific systematisations of cross-chain bridge hacks \citep{lee2023bridge,belenkov2025bridge} likewise attribute the largest bridge losses to key- and signature-level compromise rather than to contract-logic defects.

These works either evaluate tools or enumerate mechanisms within a single vulnerability class or protocol type. The present study is complementary and deliberately aggregate: it measures the joint empirical distribution of what auditors report and what attackers realise, across the full public record of both audit findings and incidents over a multi-year window, and quantifies the divergence between the two.

\paragraph{Contributions.}
(i) We assemble and classify the largest cross-firm corpus of public audit findings analysed to date (23{,}818 findings, 22 firms) alongside a quality-filtered incident dataset (218 events, US\$7.76\,B). (ii) We document that the audit-finding distribution is stable across the window while the realised-loss distribution is not, and that the two distributions are categorically misaligned. (iii) We characterise the realised-loss distribution as heavy-tailed and concentrated by chain and protocol type, with implications for risk provisioning.

The paper is organised as follows. Section~\ref{sec:data} documents the data sources and methodology. Section~\ref{sec:audit-overall} reports the overall shape of the audit-findings dataset. Section~\ref{sec:temporal} reports the temporal evolution of audit output. Section~\ref{sec:incidents} presents the incident dataset. Section~\ref{sec:divergence} contrasts the two datasets. Section~\ref{sec:human} examines human-vector attacks. Section~\ref{sec:supp} reports supplementary findings. Section~\ref{sec:conclusion} concludes.

\section{Data and methodology}
\label{sec:data}

\subsection{Audit-findings dataset}
The audit-findings dataset draws on two categories of source. The first is repository-based publication by firms that maintain public GitHub archives of PDF audit reports. The second is web-native publication through structured online channels, including vendor APIs, content-management REST endpoints, and certificate or report portals. In total, the public output of 22 independent security firms is included, spanning both publication modes. Individual firms are not identified in this paper: the analysis is conducted strictly at the aggregate level, and no firm-level results are reported.

For repository-based sources, the date on which a report was first committed to the repository is used as the report's publication date, obtained via \texttt{git log -{}-diff-filter=A}. This avoids the bias introduced when the existence of a report in a repository is conflated with its first publication; reports committed prior to the observation window are excluded even if they remain in the repository.

\subsection{Classification}
Findings are assigned to a flat taxonomy of approximately twenty categories through a three-stage procedure. The first stage applies title-level pattern matching to all findings. The second stage applies extended-context reclassification to findings that did not match in stage one, using body text extracted from the surrounding report where available. The third stage applies targeted large-language-model reclassification to remaining Critical- and High-severity findings and to the residual `Other' bucket. The `Other' rate on the full 23{,}818-finding dataset is 7.9\%. Findings originating from web-native sources classify more reliably than PDF-extracted findings because their presentation is already machine-readable; PDF-extracted findings are noisier where titles are truncated or line-broken during extraction.

\subsection{Incident dataset}
Incident-level data is sourced from \emph{rekt.news} with the publisher's explicit written permission. Each incident is tagged with a date, an approximate loss amount in US dollars, a chain, a protocol type, and a primary root cause. Loss amounts are used only where the scraped value reasonably corresponds to the dollar value of assets actually stolen; entries where the scraped amount clearly reflects total value locked or token market capitalisation rather than an actual theft are excluded, or the loss is normalised with reference to the published post-mortem. Root-cause labels are retained in our own taxonomy rather than the \emph{rekt.news} free-text field, because the two use different terminological conventions. The quality filter (a present, theft-consistent loss amount, excluding rug-pulls and editorial entries) reduces an initial archive of 242 entries to the 218 incidents analysed here, with aggregate losses of US\$7.764\,B.

\subsection{Observation window}
The window covers 1 January 2022 through 27 March 2026. The final three months represent a partial year and are labelled \emph{2026*} throughout. Year-over-year comparisons that include 2026 are restricted to those where the partial-year nature does not materially bias interpretation. Severity backfill lag---the empirical tendency of informational-severity findings to appear in public archives later than higher-severity ones---is documented in Section~\ref{sec:temporal} and treated as a known source of apparent early-year skew.

\subsection{Analytical caveat}
Public audit findings and realised exploit incidents describe different populations. The audit dataset enumerates vulnerabilities identified in code reviewed within a defined scope. The incident dataset enumerates vulnerabilities actually exploited in deployed systems, which may or may not have been audited. The two are presented in parallel to enable comparison of categorical and temporal patterns, not to imply that either is a statistical sample of the other. Dataset construction, filtering, and classification were performed with AI-assisted tooling under human editorial oversight; methodology, data-quality thresholds, and interpretation were defined and validated by the author. Analytical code and intermediate datasets are retained for reproducibility.

\section{Audit findings: overall distribution}
\label{sec:audit-overall}

\subsection{Distribution by severity}
\begin{figure}[t]
\centering
\includegraphics[width=0.78\linewidth]{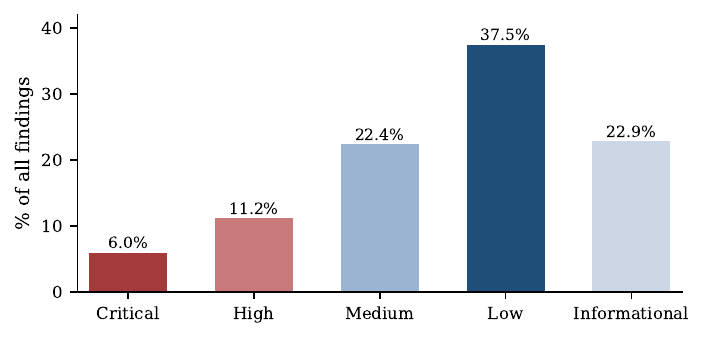}
\caption{Distribution of severity classifications across all 23{,}818 findings in the observation window.}
\label{fig:severity}
\end{figure}

Across the full dataset, 17.2\% of findings are classified as Critical or High, comprising 1{,}439 Critical and 2{,}659 High findings. Medium-severity findings constitute 22.4\%; Low-severity 37.5\%; Informational 22.9\% (Figure~\ref{fig:severity}). The large share of Low and Informational findings is consistent with industry practice, in which audit reports function as holistic code-quality assessments rather than narrowly focused vulnerability enumerations.

\subsection{Distribution by vulnerability category}
Table~\ref{tab:cats} reports the frequency of vulnerability categories. The five most frequent---logic/business-logic, code quality, input validation, access control, and initialisation/upgradeability---together account for 54.7\% of all findings. The distribution has a long tail: the top sixteen categories account for 92.1\% of the dataset, with the remaining 7.9\% in the `Other' residual. Three observations are worth noting. First, reentrancy---historically emblematic of smart-contract vulnerability---accounts for only 2.1\% of findings, indicating that auditors now surface this class substantially less frequently than in the mid-2010s. Second, logic and business-logic errors constitute the single largest category, reflecting that many findings concern protocol-specific invariant violations rather than named CWE patterns. Third, oracle and price-manipulation findings account for only 4.3\% of audit output but, as Section~\ref{sec:divergence} reports, are associated with a substantially larger share of realised losses.

\begin{table}[t]
\centering
\caption{Category frequencies across all 23{,}818 findings, 2022--Q1 2026.}
\label{tab:cats}
\small
\begin{tabular}{l r r}
\toprule
Category & Count & Percent \\
\midrule
Logic Error / Business Logic & 3479 & 14.6 \\
Code Quality & 3098 & 13.0 \\
Input Validation & 2372 & 10.0 \\
Access Control / Authorization & 2339 & 9.8 \\
Other & 1882 & 7.9 \\
Initialization / Upgradeability & 1749 & 7.3 \\
Integer Overflow / Arithmetic & 1278 & 5.4 \\
Oracle / Price Manipulation & 1016 & 4.3 \\
Gas / Efficiency & 992 & 4.2 \\
Cross-chain / Bridge & 881 & 3.7 \\
Signature / Replay Attack & 797 & 3.3 \\
Denial of Service & 781 & 3.3 \\
Precision / Rounding Errors & 562 & 2.4 \\
Token Standard Compliance & 544 & 2.3 \\
Unchecked Return Values & 496 & 2.1 \\
Reentrancy & 491 & 2.1 \\
\bottomrule
\end{tabular}
\end{table}

\subsection{Distribution by technology stack}
\begin{figure}[t]
\centering
\includegraphics[width=\linewidth]{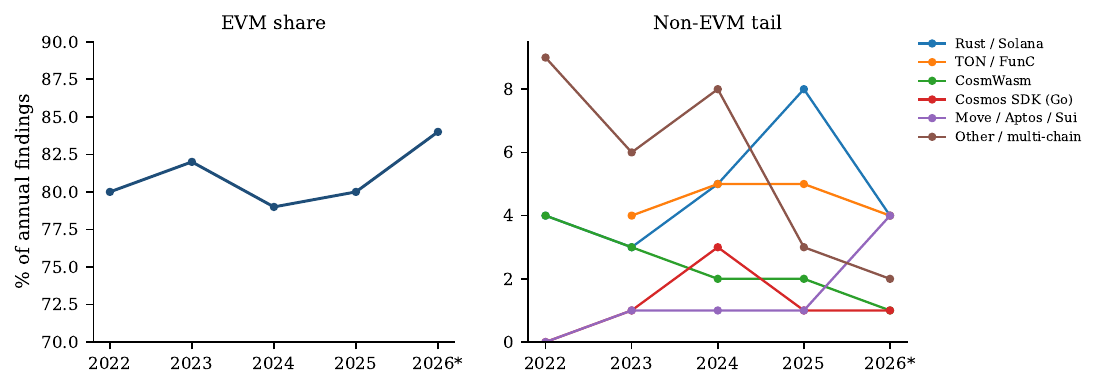}
\caption{Distribution of audit findings across technology stacks, as percentage of annual findings. Left: dominant EVM share. Right: composition of the non-EVM tail.}
\label{fig:stack}
\end{figure}

Solidity and EVM-compatible chains generate between 79\% and 84\% of findings in every year (Figure~\ref{fig:stack}). The residual 16--21\% is distributed across a changing composition of non-EVM stacks. Rust/Solana rises from approximately 3\% in 2023 to a local peak of 8\% in 2025 before retreating to 4\% in Q1~2026. TON/FunC does not appear as a distinct category before 2023 and stabilises thereafter at 4--5\%. CosmWasm and Cosmos~SDK combined range between 5\% and 8\%, with Cosmos~SDK becoming more visible from 2024 as chain-module audits entered public circulation. Move-based ecosystems (Aptos, Sui) constitute a smaller but fastest-growing segment in proportional terms, reaching 4\% of Q1~2026 findings.

\section{Temporal evolution, 2022--Q1 2026}
\label{sec:temporal}

\subsection{Volume and severity}
\begin{figure}[t]
\centering
\begin{minipage}{0.49\linewidth}\centering
\includegraphics[width=\linewidth]{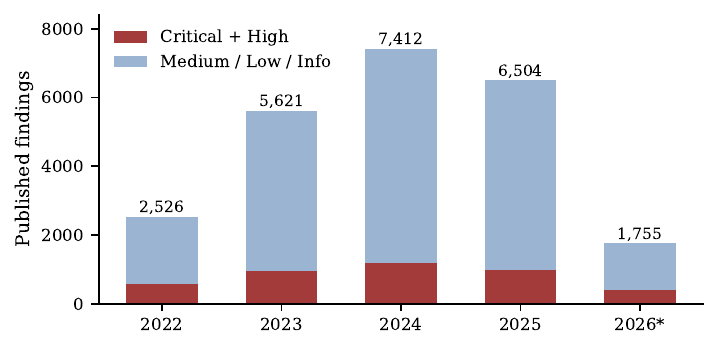}
\end{minipage}\hfill
\begin{minipage}{0.49\linewidth}\centering
\includegraphics[width=\linewidth]{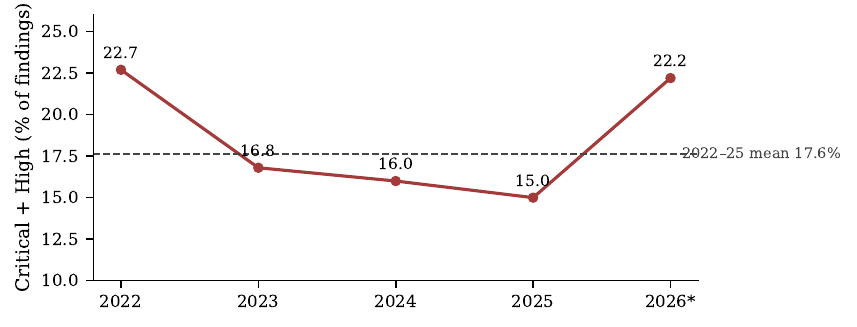}
\end{minipage}
\caption{Left: annual count of published findings, decomposed into Critical+High versus all other severities (2023 total inferred as the residual of the annual totals). Right: share of Critical+High findings per year, with the 2022--2025 mean shown for reference.}
\label{fig:volume}
\end{figure}

Published audit volume more than doubled between 2022 and 2024, rising from 2{,}526 findings to 7{,}412 (Figure~\ref{fig:volume}). Volume retreated to 6{,}504 in 2025 and is running at 1{,}755 findings for Q1~2026, a rate consistent with the 2023 pace on an annualised basis. Growth from 2022 through 2024 reflects, in substantial part, the entry of new firms into public-report publication (twelve firms in 2022, seventeen in 2024) rather than increased identification rates per report. The 2025 contraction is consistent with a softer audit market.

The Critical-plus-High share is remarkably stable across complete years: 22.7\% in 2022, 16.8\% in 2023, 16.0\% in 2024, and 15.0\% in 2025. The elevated 22.2\% reading for Q1~2026 is partially attributable to the empirical pattern that Informational and Low-severity findings are backfilled to public archives with greater lag than higher-severity findings; early-year readings therefore over-represent the serious categories. Over the four complete years, the Critical-plus-High share is essentially flat.

\subsection{Category evolution}
\begin{figure}[t]
\centering
\includegraphics[width=0.62\linewidth]{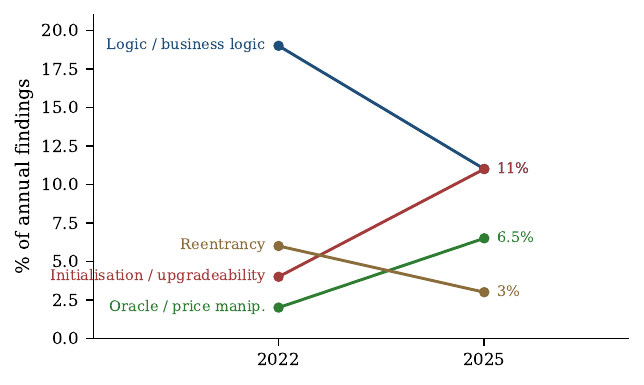}
\caption{Movement in the share of audit findings for four categories between 2022 and 2025 (stated endpoints).}
\label{fig:catslope}
\end{figure}

The categorical composition exhibits several distinct movements within an otherwise stable overall distribution (Figure~\ref{fig:catslope}). The share of logic and business-logic findings declined from approximately 19\% in 2022 to approximately 11\% in 2025--2026; a substantial portion of this decline is reclassification-driven, as the pipeline routes findings that earlier datasets captured under a broad `logic' label into more specific categories. The share of initialisation and upgradeability findings rose from approximately 4\% to approximately 11\%, consistent with the industry-wide shift from monolithic contract architectures to proxy-upgradeable patterns and the vulnerability classes associated with initialisation, admin-role management, and storage-slot collisions. The share of oracle and price-manipulation findings more than doubled, from approximately 2\% to 6--7\%. The share of reentrancy findings declined further, from approximately 6\% to approximately 3\%; Section~\ref{sec:incidents} reports that the residual reentrancy bugs reaching production remain associated with disproportionate realised losses. Cross-chain and bridge findings peaked in 2023 and have trended downward, consistent with reduced novel-bridge launch rates and increased audit coverage of the existing bridge generation.

\subsection{Stability alongside movement}
Against the categorical drift documented above, three features of the dataset are essentially unchanged over the complete four-year period: the Critical-plus-High share (within a 15--17\% band); the identity of the top five categories (access control, logic, initialisation, input validation, and arithmetic, with some reordering); and the dominance of Solidity/EVM (79--84\% of annual findings). The audit picture does not reconfigure itself year-over-year; it is stable at a level consistent with an industry that has developed its review practices to address a known set of recurring failure modes.

\subsection{Technology-stack shifts}
Table~\ref{tab:stack} presents the year-over-year evolution of the technology-stack distribution. The EVM share is stable; the composition of the non-EVM tail is not. TON/FunC is absent from 2022 public audit output because reporting conventions for that stack crystallised in mid-2023. The 2025 Rust/Solana peak coincides with a renewed phase of Solana DeFi activity. The Move ecosystem exhibits the fastest proportional growth, though from a small absolute base.

\begin{table}[t]
\centering
\caption{Technology-stack distribution of audit findings, percentage of annual findings.}
\label{tab:stack}
\small
\begin{tabular}{l ccccc}
\toprule
Stack & 2022 & 2023 & 2024 & 2025 & 2026* \\
\midrule
Solidity / EVM & 80\% & 82\% & 79\% & 80\% & 84\% \\
Rust / Solana & 4\% & 3\% & 5\% & 8\% & 4\% \\
TON / FunC & --- & 4\% & 5\% & 5\% & 4\% \\
CosmWasm & 4\% & 3\% & 2\% & 2\% & 1\% \\
Cosmos SDK (Go) & 0\% & 1\% & 3\% & 1\% & 1\% \\
Move / Aptos / Sui & 0\% & 1\% & 1\% & 1\% & 4\% \\
Other / multi-chain & 9\% & 6\% & 8\% & 3\% & 2\% \\
\bottomrule
\end{tabular}
\end{table}

\section{Real-world exploit incidents}
\label{sec:incidents}
Over the window, the \emph{rekt.news} archive documents 218 incidents meeting the quality filters of Section~\ref{sec:data}, with aggregate losses of approximately US\$7.76\,B. Table~\ref{tab:annual} summarises the annual decomposition. Losses are volatile year-over-year because single large events dominate annual totals: 2022 is influenced heavily by the Ronin Network exploit (US\$624\,M); 2025 by the Bybit multi-signature phishing incident (US\$1.44\,B); the 2025 DeFi-only loss ranking is headed by the Cetus arithmetic-overflow exploit on Sui (US\$223\,M).

\begin{figure}[t]
\centering
\includegraphics[width=0.72\linewidth]{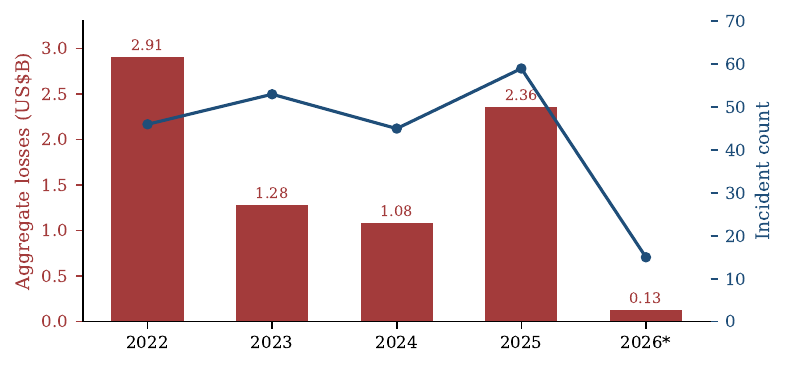}
\caption{Documented exploit incidents (line, right axis) and aggregate annual losses (bars, left axis), 2022--Q1 2026.}
\label{fig:incidents}
\end{figure}

\begin{table}[t]
\centering
\caption{Annual decomposition of exploit activity. The `Audited' column reports the count of incidents where the affected protocol had received at least one public audit.}
\label{tab:annual}
\small
\begin{tabular}{l c c c c}
\toprule
Year & Incidents & Total losses & Audited & Audited losses \\
\midrule
2022 & 46 & US\$2.91\,B & 23 & $\approx$ US\$1.42\,B \\
2023 & 53 & US\$1.28\,B & 37 & $\approx$ US\$820\,M \\
2024 & 45 & US\$1.08\,B & 23 & $\approx$ US\$350\,M \\
2025 & 59 & US\$2.36\,B & 17 & $\approx$ US\$1.67\,B \\
2026* & 15 & US\$0.13\,B & 5 & $\approx$ US\$40\,M \\
\midrule
Full window & 218 & US\$7.76\,B & 105 & $\approx$ US\$4.30\,B \\
\bottomrule
\end{tabular}
\end{table}

\subsection{Root-cause distribution}
Table~\ref{tab:rootcause} reports the distribution of aggregate losses by root cause, ordered by total dollar impact. Three categories dominate. Private-key compromise alone accounts for US\$1{,}894\,M (24.4\% of cumulative losses). Phishing and social engineering account for US\$1{,}511\,M (19.5\%). Access-control failures account for US\$994\,M (12.8\%). Together these three represent 56.7\% of aggregate realised losses. Private-key compromise combines a relatively high incidence count (45) with a high mean loss ($\approx$US\$42\,M) to produce the largest aggregate. Phishing and social-engineering incidents are few (12) but exhibit the second-highest mean loss ($\approx$US\$126\,M), disproportionately influenced by the single Bybit event. Bridge exploits and signature/replay attacks are rare but exhibit the highest mean per-incident losses of any category: the four bridge events average US\$148\,M each, and the two signature-replay events---Wormhole (February 2022) and Nomad (August 2022)---average US\$258\,M.

\begin{table}[t]
\centering
\caption{Distribution of aggregate realised losses by root-cause category, full observation window.}
\label{tab:rootcause}
\small
\begin{tabular}{l c c c c}
\toprule
Root cause & Incidents & Total losses & \% losses & Mean / incident \\
\midrule
Private Key Compromise & 45 & US\$1{,}894\,M & 24.4 & US\$42.1\,M \\
Phishing / Social Engineering & 12 & US\$1{,}511\,M & 19.5 & US\$125.9\,M \\
Access Control & 30 & US\$994\,M & 12.8 & US\$33.1\,M \\
Oracle / Price Manipulation & 43 & US\$666\,M & 8.6 & US\$15.5\,M \\
Bridge Exploit & 4 & US\$593\,M & 7.6 & US\$148.3\,M \\
Signature / Replay Attack & 2 & US\$516\,M & 6.6 & US\$258.0\,M \\
Logic Error / Business Logic & 37 & US\$298\,M & 3.8 & US\$8.1\,M \\
Integer Overflow / Arithmetic & 7 & US\$288\,M & 3.7 & US\$41.1\,M \\
Flash Loan & 5 & US\$266\,M & 3.4 & US\$53.1\,M \\
Reentrancy & 14 & US\$256\,M & 3.3 & US\$18.3\,M \\
Supply Chain / Dependency & 7 & US\$243\,M & 3.1 & US\$34.7\,M \\
Governance Attack & 5 & US\$206\,M & 2.7 & US\$41.2\,M \\
\bottomrule
\end{tabular}
\end{table}

\section{Divergence between audit output and exploit activity}
\label{sec:divergence}
\begin{figure}[t]
\centering
\includegraphics[width=\linewidth]{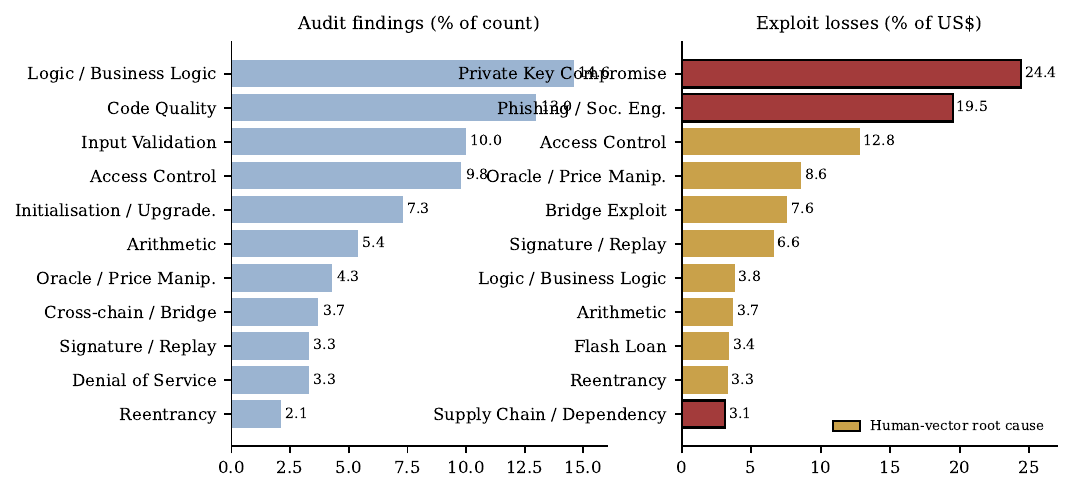}
\caption{Left: audit findings by category (\% of count, 2022--Q1 2026). Right: realised exploit losses by root cause (\% of US\$). Human-vector root causes on the right panel are outlined in black.}
\label{fig:divergence}
\end{figure}

The central empirical finding is that the categorical distribution of published audit findings does not correspond to the categorical distribution of realised exploit losses. Figure~\ref{fig:divergence} presents the two distributions side by side.

\subsection{Observed misalignment}
Of the twelve most frequent audit categories and the twelve largest exploit-loss root causes, access control is the only category appearing in the top four on both sides (Table~\ref{tab:rank}). The remaining top-four audit categories---logic errors, code quality, and input validation---together account for 37.6\% of audit output but only 3.8\% of realised dollar losses (predominantly through the Logic Error category, in position seven on the loss side). Conversely, private-key compromise, phishing and social engineering, and supply-chain dependency attacks together account for approximately 47\% of realised losses but constitute a negligible share of audit findings.

\begin{table}[t]
\centering
\caption{Rank comparison of audit-finding categories against exploit-loss root causes.}
\label{tab:rank}
\small
\begin{tabular}{p{0.46\linewidth} p{0.46\linewidth}}
\toprule
Rank by audit frequency & Rank by exploit losses \\
\midrule
1. Logic Error / Business Logic (14.6\%) & 1. Private Key Compromise (US\$1.89\,B) \\
2. Code Quality (13.0\%) & 2. Phishing / Social Engineering (US\$1.51\,B) \\
3. Input Validation (10.0\%) & 3. Access Control (US\$994\,M) \\
4. Access Control / Authorization (9.8\%) & 4. Oracle / Price Manipulation (US\$666\,M) \\
5. Initialization / Upgradeability (7.3\%) & 5. Bridge Exploit (US\$593\,M) \\
6. Integer Overflow / Arithmetic (5.4\%) & 6. Signature / Replay Attack (US\$516\,M) \\
7. Oracle / Price Manipulation (4.3\%) & 7. Logic Error / Business Logic (US\$298\,M) \\
8. Reentrancy (2.1\%) & 8. Integer Overflow / Arithmetic (US\$288\,M) \\
9. Denial of Service (3.3\%) & 9. Reentrancy (US\$256\,M) \\
10. Cross-chain / Bridge (3.7\%) & 10. Supply Chain / Dependency (US\$243\,M) \\
\bottomrule
\end{tabular}
\end{table}

\subsection{Interpretation}
The misalignment admits a specific interpretation. A conventional smart-contract audit reviews the source code of a defined commit against a defined scope. It does not review the deployment environment, signer key-management practices, the continuous-integration pipeline, front-end hosting infrastructure, or the supply chain of third-party dependencies. The categories that audits surface are therefore, by construction, those visible in source code under review. The categories that drive realised losses include several classes of attack---principally operational-security failures and human-facing manipulation---that are not visible in source code, whether because they do not exist there (key theft, phishing) or because they arise in artefacts outside the audited scope (supply-chain compromise, governance-vote manipulation after a discrete contract has been audited).

This should not be read as a claim that audits are ineffective. The relevant conclusion is more specific: the effective scope of a conventional audit is narrower than the risk surface that the word `security' is implicitly taken to cover in the Web3 context. Access control is the exception that illustrates the rule, because it sits at the interface between source code and operational reality, and is therefore the one category where a static code review and a deployed-system attack can meaningfully meet. This aggregate divergence is consistent with prior findings at the level of individual contracts: that flagged vulnerabilities are rarely exploited in practice \citep{perez2021vulnerable}, and that contemporary analysis tools would have prevented only a minority of high-impact real-world attacks \citep{chaliasos2024tools}.

\section{The prominence of human-vector attacks}
\label{sec:human}
We classify four root-cause categories as \emph{human-vector}: private-key compromise, phishing and social engineering, supply-chain or dependency compromise, and governance attack. The defining feature common to this grouping is that the proximate cause of the realised loss sits outside the static source code of the affected protocol---whether at the operational-security layer (keys, signer workflows), the distribution layer (front-ends, build artefacts, third-party packages), or the governance layer (vote mechanisms operating after deployment).

\subsection{Temporal pattern}
\begin{figure}[t]
\centering
\includegraphics[width=0.72\linewidth]{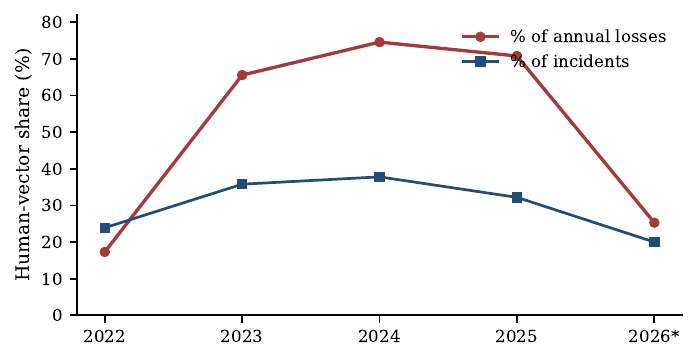}
\caption{Share of annual losses and of incident counts attributable to human-vector root causes.}
\label{fig:human}
\end{figure}

Human-vector attacks accounted for the majority of annual losses in each of 2023, 2024, and 2025, with loss shares of 65.6\%, 74.6\%, and 70.8\% respectively (Table~\ref{tab:human}, Figure~\ref{fig:human}). The 2022 pattern is the exception: that year's losses concentrated in a small number of large bridge exploits (Ronin, Wormhole, Nomad, Harmony, BNB Bridge) whose root causes were either contract-level or signature-level rather than operational.

\begin{table}[t]
\centering
\caption{Annual incidence and loss share of human-vector attacks.}
\label{tab:human}
\small
\begin{tabular}{l c c c c c}
\toprule
Year & Incidents & Human-vector & \% of incidents & H-V losses & \% of losses \\
\midrule
2022 & 46 & 11 & 23.9 & US\$502\,M & 17.3 \\
2023 & 53 & 19 & 35.8 & US\$843\,M & 65.6 \\
2024 & 45 & 17 & 37.8 & US\$803\,M & 74.6 \\
2025 & 59 & 19 & 32.2 & US\$1{,}667\,M & 70.8 \\
2026* & 15 & 3 & 20.0 & US\$34\,M & 25.3 \\
\midrule
Full window & 218 & 69 & 31.7 & US\$3{,}849\,M & 49.6 \\
\bottomrule
\end{tabular}
\end{table}

\subsection{Illustrative incidents}
A small number of large incidents define the human-vector pattern. \textbf{Bybit} (February 2025, US\$1.44\,B): a multi-signature approval workflow was compromised through targeted phishing; signers approved a disguised upgrade of the cold-wallet contract. The underlying contract had been audited; the approval interface had not. This is the largest single cryptocurrency theft to date and accounts for 18.4\% of cumulative losses in the window. \textbf{DMM Bitcoin} (May 2024, US\$304\,M): private keys stolen, no audit-visible code defect. \textbf{WazirX} (July 2024, US\$235\,M): multi-signature workflow compromise. \textbf{Mixin Network} (September 2023, US\$200\,M): supply-chain compromise of cloud-hosted database credentials. \textbf{Ronin Network} (March 2022, US\$624\,M): validator-key compromise of five of nine validators, not a contract-layer event. \textbf{Poloniex} (November 2023, US\$126\,M): hot-wallet private-key compromise at an exchange.

\subsection{The audited-but-exploited pattern}
Of the 218 incidents, 105 (48\%) affected protocols that had received at least one public audit prior to the event. These account for approximately US\$4.3\,B of the US\$7.76\,B aggregate, or roughly 55\% of cumulative losses. Table~\ref{tab:audited} lists the twelve largest such incidents. Of these twelve, nine fall into root-cause categories outside the conventional smart-contract audit scope (phishing, private-key compromise, signature replay against infrastructure rather than the audited contract, supply-chain compromise, or governance-mechanism manipulation). The remaining three---Nomad (initialisation), Euler (donation attack), Qubit (logic error)---are cases where the audit-scope root cause was real but the affected version was either post-audit, outside the explicit scope, or arose in a path not fully enumerated during review.

\begin{table}[t]
\centering
\caption{Largest exploit incidents affecting protocols with prior public audit coverage. Audit attribution is drawn from the \emph{rekt.news} archive.}
\label{tab:audited}
\small
\begin{tabular}{l c c l}
\toprule
Protocol & Date & Loss & Root cause \\
\midrule
Bybit & 2025-02-22 & US\$1{,}430\,M & Phishing of multi-sig signers \\
BNB Bridge & 2022-10-07 & US\$586\,M & Signature replay on IAVL Merkle proofs \\
Mixin Network & 2023-09-25 & US\$200\,M & Supply-chain (cloud DB credentials) \\
Euler Finance & 2023-03-14 & US\$197\,M & Flash-loan donation attack \\
Nomad Bridge & 2022-08-02 & US\$190\,M & Initialisation bug in signature check \\
Beanstalk & 2022-04-18 & US\$181\,M & Governance attack (flash-loaned votes) \\
Poloniex & 2023-11-10 & US\$126\,M & Private-key compromise \\
Harmony Bridge & 2022-06-24 & US\$100\,M & Private-key compromise (2 of 5 signers) \\
Heco / HTX & 2023-11-22 & US\$99\,M & Private-key compromise \\
Orbit Bridge & 2024-01-03 & US\$82\,M & Private-key compromise \\
FEI / Rari & 2022-05-01 & US\$80\,M & Reentrancy \\
Qubit Finance & 2022-01-28 & US\$80\,M & Logic error in bridge deposit \\
\bottomrule
\end{tabular}
\end{table}

\section{Supplementary findings}
\label{sec:supp}

\subsection{Concentration of losses (Pareto)}
\begin{figure}[t]
\centering
\includegraphics[width=0.70\linewidth]{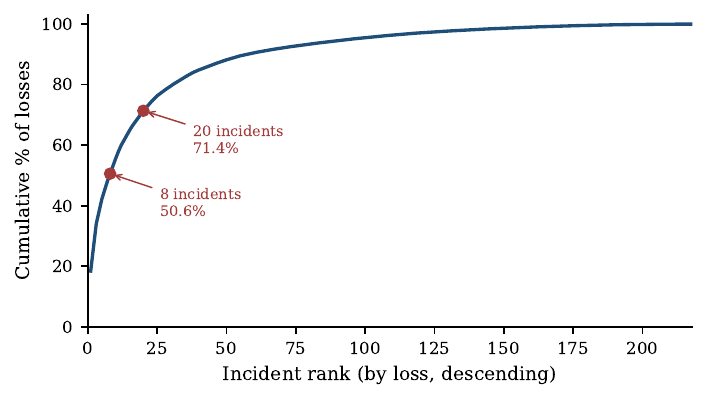}
\caption{Cumulative share of aggregate losses as a function of incident rank.}
\label{fig:pareto}
\end{figure}

Aggregate losses exhibit a steeply concentrated distribution (Figure~\ref{fig:pareto}). The single largest incident (Bybit) alone accounts for 18.4\% of cumulative losses. The eight largest account for 50.6\%; the twenty largest for 71.4\%. The remaining 198 incidents---91\% of the count---together account for less than 29\% of cumulative losses. The distributional shape is inconsistent with an assumption of approximately Gaussian losses; risk models assuming Gaussianity will systematically underestimate annual worst-case outcomes.

\subsection{Chain concentration}
\begin{figure}[t]
\centering
\includegraphics[width=0.78\linewidth]{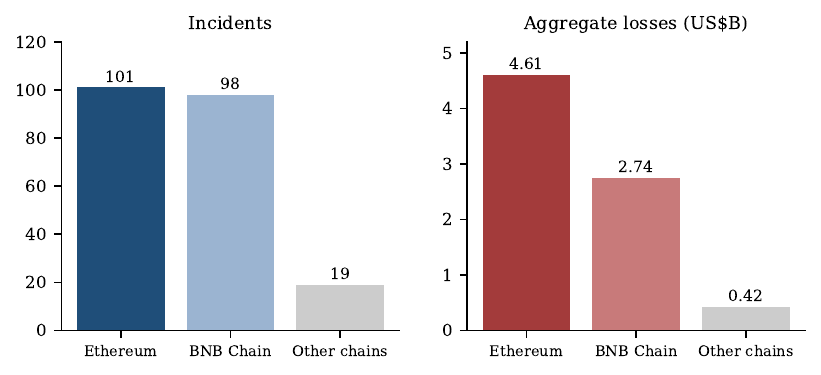}
\caption{Incident counts and aggregate losses by chain (BNB Chain aggregates BSC and BNB Beacon Chain).}
\label{fig:chain}
\end{figure}

Incident activity is concentrated by chain to an even greater degree than by root cause (Figure~\ref{fig:chain}). Ethereum and BNB Chain together host 89\% of all incidents (199 of 218) and 94\% of all losses (US\$7.35\,B of US\$7.76\,B). Every other chain---Solana, Sui, Arbitrum, Stacks, Stellar, Cosmos, Base, Sonic, and several smaller chains---contributes a single-digit incident count across the full window. The pattern admits two interpretations: either chains outside the top two have held up well under adversarial pressure, or most do not yet host sufficient total value locked to justify large-scale attacker investment, and incident concentration will broaden as they grow. The 2025 Cetus exploit on Sui (US\$223\,M) is the first substantive indication that the second interpretation is in part correct: a single large loss can shift the risk profile of a chain materially.

\subsection{Distributional shape by root cause}
\begin{figure}[t]
\centering
\includegraphics[width=0.62\linewidth]{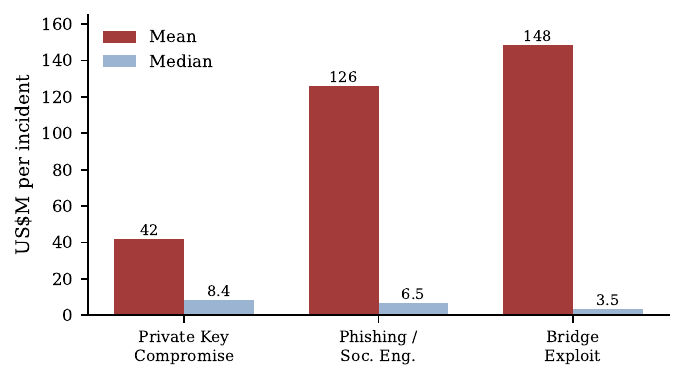}
\caption{Mean versus median loss per incident, for three root causes with $n\geq 3$ incidents (median for private-key compromise inferred from the reported mean-to-median ratio).}
\label{fig:meanmedian}
\end{figure}

Because the loss distribution is heavy-tailed, mean per-incident values are unreliable summary statistics for categories that include one or more catastrophic events (Figure~\ref{fig:meanmedian}). The mean phishing incident is US\$126\,M; the median is US\$6.5\,M, a ratio of approximately 19. The mean bridge exploit is US\$148\,M; the median is US\$3.5\,M, a ratio of approximately 42, reflecting the outsized 2022 events. Even private-key compromise, the most uniformly large human-vector category, exhibits a mean-to-median ratio of approximately 5. The practical implication is that planning against a category of attack should be conducted against the tail of the distribution rather than its mean. For centralised-exchange and bridge risk in particular, the mean per-incident loss substantially understates the distribution of plausible outcomes.

\subsection{Protocol-type concentration of losses}
Centralised exchanges and bridges together account for approximately 60\% of cumulative losses across 21\% of incidents (Table~\ref{tab:ptype}). Lending, the most numerous protocol type by incident count, accounts for less than 10\% of cumulative losses. Protocol type thus functions as a stronger predictor of tail-loss exposure than the specific vulnerability class most commonly associated with that type.

\begin{table}[t]
\centering
\caption{Distribution of aggregate losses by protocol type.}
\label{tab:ptype}
\small
\resizebox{\linewidth}{!}{%
\begin{tabular}{l c c c l}
\toprule
Protocol type & Incidents & Total losses & Mean / incident & Typical failure mode \\
\midrule
CEX (centralised exchange) & 24 & US\$2{,}284\,M & US\$95.2\,M & Key compromise, phishing \\
Bridge & 22 & US\$2{,}401\,M & US\$109.1\,M & Signature/replay, key compromise \\
Lending & 48 & US\$736\,M & US\$15.3\,M & Oracle manipulation, logic \\
DEX & 31 & US\$603\,M & US\$19.5\,M & Arithmetic, oracle, reentrancy \\
Wallet / custody & 7 & US\$400\,M & US\$57.2\,M & Key compromise \\
Derivatives & 11 & US\$71\,M & US\$6.5\,M & Oracle, logic \\
Yield / staking & 8 & US\$40\,M & US\$5.0\,M & Logic, reentrancy \\
DAO & 3 & US\$6\,M & US\$2.0\,M & Governance \\
\bottomrule
\end{tabular}}
\end{table}

\subsection{The audit-coverage paradox, reconsidered}
A na\"ive reading of Table~\ref{tab:annual}---48\% of incidents affected audited protocols, accounting for 55\% of losses---might support the conclusion that audits do not reduce realised losses. The data does not support that conclusion. The majority of audited-but-exploited incidents have root causes outside the audited scope, as Section~\ref{sec:human} documents. Where a contract-level bug was the direct cause (Nomad initialisation, Euler donation flash-loan, and others), the audit and the deployed vulnerable code diverged in identifiable ways: subsequent code changes, explicitly out-of-scope paths, or undeployed audited versions. A more accurate reading is that audits are narrowly effective but broadly incomplete relative to the risk surface they are often implicitly relied upon to defend. The category of protocol whose failure mode is primarily operational---custodians, bridges, exchanges---is the same category for which a static contract review provides the least defensive coverage and for which complementary operational-security engagements are most impactful on the loss distribution.

\section{Conclusion}
\label{sec:conclusion}
The analysis supports five empirical claims across the observation window. (1)~The distribution of public audit findings is substantively stable: the Critical-plus-High share has remained within a narrow band (15--17\% in complete years), and the identity of the five most frequent categories has not changed; drift is present at the individual-category level---most notably the rise of initialisation and oracle findings, and the decline of reentrancy---but the overall distribution does not exhibit regime change. (2)~The distribution of realised exploit losses has shifted substantially: from 2023 onwards, human-vector attacks account for the majority of annual losses, in contrast to the comparatively stable code-review picture described by claim~(1). (3)~Audit output and realised exploit output describe different populations whose categorical distributions do not correspond; access control is the only category in the top four of both, and the misalignment substantially reflects the narrower scope of a conventional audit (source code of a specific commit) relative to the broader risk surface on which losses are realised (operational security, deployment infrastructure, and human-facing approval workflows). (4)~The distribution of realised losses is heavy-tailed: eight incidents account for 50.6\% of cumulative losses and twenty for 71.4\%; chain-level and protocol-type concentration (89\% of incidents on two chains; 60\% of losses in two protocol types) reinforces the tail dependence, and security programmes designed against mean outcomes systematically under-provision against tail outcomes. (5)~Solidity and EVM remain the dominant stack throughout, but the long tail of non-EVM stacks is present and growing, and teams designing security programmes for multi-stack products on the assumption of single-stack audit availability are accumulating coverage gaps the public-report record does not yet fully illuminate.

The overall picture does not support a claim that the ecosystem has become more or less secure in any simple sense. The code-review problem, as surfaced by public audit output, is approximately where it was in 2022. The operational-security problem, as surfaced by the incident record, grew substantially through 2023--2025. Both components require continued attention, from disciplines that have historically operated with relatively little overlap. The data supports a portfolio view of Web3 security practice in which code review and operational-security engagements are treated as complementary rather than substitutable: audits for the bugs that exist in the code as written, and parallel assessments---of key management, signer workflows, build-pipeline hardening, and dependency supply chain---for the categories that drive the empirically larger share of realised losses.

\section*{Data and reproducibility}
Incident-level metadata is licensed from \emph{rekt.news} under explicit written permission for the factual fields used here (dates, loss amounts, chain, protocol, and root-cause classification); no article text is reproduced. All \emph{rekt.news} content remains the copyright of Rekt News (EU trademark registration 018857408). Analytical code and intermediate datasets are retained and available on request.

\section*{Acknowledgements}
The author gratefully acknowledges \emph{rekt.news} for granting explicit written permission to use their incident archive as a primary data source; the real-world exploit analysis in this paper would not have been possible without that collaboration. The author also thanks the independent security firms whose public audit output forms the basis of the findings dataset. Public disclosure of audit results---whether through repository-based publication or web-native report portals---is what makes empirical work of this kind possible. Any errors in the aggregation, classification, or interpretation of the data are the author's alone.

\end{document}